\documentclass[runningheads,a4paper]{llncs}

\usepackage{amssymb}
\usepackage{graphicx}

\usepackage{siunitx}
\usepackage{amsmath}
\usepackage{comment}

\usepackage{url}   
\newcommand{\keywords}[1]{\par\addvspace\baselineskip
\noindent\keywordname\enspace\ignorespaces#1}

\begin{document}

\mainmatter  

\title{State-space analysis of an Ising model reveals contributions of pairwise interactions to sparseness, fluctuation, and stimulus coding of monkey V1 neurons}

\titlerunning{Neural interactions contribute to sparseness, fluctuation, and stimulus coding}

%
%
\author{Jimmy Gaudreault\inst{1} \and Hideaki Shimazaki\inst{2,3}}
\authorrunning{Neural interactions contribute to sparseness, fluctuation, and stimulus coding}

\institute{Polytechnique Montreal, Quebec, Canada                   \and
Graduate School of Informatics, Kyoto University 
\and 
Honda Research Institute Japan\\
\url{jimmy.gaudreault@polymtl.ca, }
\url{h.shimazaki@kyoto-u.ac.jp}}

%
%

\toctitle{Lecture Notes in Computer Science}
\tocauthor{Ising model to analyze contribution of interactions in monkey V1 neurons}
\maketitle

\begin{abstract}
\emph{In this study, we analyzed the activity of monkey V1 neurons responding to grating stimuli of different orientations using inference methods for a time-dependent Ising model. The method provides optimal estimation of time-dependent neural interactions with credible intervals according to the sequential Bayes estimation algorithm. Furthermore, it allows us to trace dynamics of macroscopic network properties such as entropy, sparseness, and fluctuation. Here we report that, in all examined stimulus conditions, pairwise interactions contribute to increasing sparseness and fluctuation. We then demonstrate that the orientation of the grating stimulus is in part encoded in the pairwise interactions of the neural populations. These results demonstrate the utility of the state-space Ising model in assessing contributions of neural interactions during stimulus processing.}

\keywords{Neural Interactions, Neural Coding, Macroscopic Network Properties, Bayesian Inference, Binary Time-series}
\end{abstract}

\section{Introduction}
Since neural population activity is constrained by external stimuli and biophysical mechanisms of the neural networks, understanding the statistical regularity of the population activity is an important step toward revealing these underlying mechanisms and further elucidating stimulus coding strategies by the populations of neurons. In order to understand their complex activity patterns, an Ising model has been applied frequently (see \cite{Schneidman2006,Ganmor2011,Tkaik2014} and references therein). This model originally developed in statistical mechanics to describe interacting magnetic spins is suitable for analyzing the collective behavior of binary patterns. It is also used in machine learning applications as the Botlzmann machine. 

Most of the analyses using the Ising model assumed stationary data in which firing rates and correlations are expected to be constant in time. The static model prohibited analyses of in-vivo data, in which firing rates and even correlations are known to evolve over time \cite{Aertsen89,Vaadia95}. As a solution, a state-space model was developed that augmented the stationary Ising model to one that considers dynamics in both firing rates and correlations \cite{Shimazaki2009,Shimazaki2012,Donner2017}. However, the utility of the method has not been fully demonstrated yet. 

In this study, we analyzed the activity of V1 neurons using the state-space Ising model. We report that pairwise interactions contribute to increasing temporal sparseness and fluctuation, and encoding stimulus information. 

\section{Methods}
\subsection{Data Description and Preprocessing} Population activity of V1 neurons of 3 anesthetized macaque monkeys exposed to visual stimulus was analyzed. It was recorded by Smith and Kohn \cite{smith2008spatial}. The data is available at CRCNS.org \cite{kohn2016utah}. The experimental methods used to perform recordings are briefly explained in \cite{smith2008spatial} and are detailed in \cite{Cavanaugh2002}. To summarize, an array of 100 microelectrodes was used to perform simultaneous recordings of approximately 100 neurons per monkey. The electrodes were implanted in the primary visual area (V1). The stimuli shown to the monkeys consisted of sinusoidal gratings at 12 different equally separated orientations from \ang{0} (vertical gratings) to \ang{330}. The spike data for each trial lasted 1.28 s. During a trial, a monkey was shown gratings of only one orientation. An isoluminant gray screen was presented during 1.5 s between trials. Temporal and spacial frequencies of the gratings were set to those typically preferred by parafoveal V1 neurons. The experiment was repeated 200 times for every stimulus orientation and for every monkey. 

The timing of spikes of different single neurons in this data set was obtained by spike sorting based on a mixture decomposition method \cite{Shoham2003}, allowing to discriminate waveforms from different neurons simultaneously measured by the microelectrodes. To consider only recordings of good quality, we excluded neurons with a signal-to-noise ratio lower than 2.75 and neurons with a firing rate lower than 2 spikes/s for all stimuli, as suggested by Smith and Kohn \cite{smith2008spatial}. This left approximately 40 neurons per monkey. 

In the present study, we analyzed nearly simultaneous activity of the neural populations. For this goal, we constructed binary spike trains by binning the spike timing sequences. Time bins ($\Delta t$) of 10 ms were used, giving a total of 128 time bins ($T$) for the duration of the stimulus presentation. For a given trial and neuron, if one or more spikes occurred between times $(i-1)\Delta t$ and $i\Delta t$ s, the value 1 is attributed to the $i^{th}$ time bin. Otherwise, the value 0 is attributed.

\subsection{The State-Space Ising Model for a Neural Population} The model used to analyze neural activity is the Ising model (or the Boltzmann machine), a model frequently used in statistical physics and machine learning. For a binary vector of length $N$, the Ising model is a probability distribution of all $2^N$ possible patterns. By considering up to pairwise interactions, the Ising model is given by
\begin{equation}
p(x_1,x_2,\hdots,x_N \vert \boldsymbol\theta) = \exp \left[\sum_{i}\theta_i x_i + \sum_{i<j}\theta_{ij} x_i x_j - \psi(\boldsymbol\theta)\right].
\label{eq:ising_model}
\end{equation}
For a neural system, $N$ is the number of neurons and the binary vector $\mathbf{x}=(x_1,x_2,\hdots,x_N)^{\prime}$ is the activity of the population, where each binary variable $x_i$ is the activity of the $i^{th}$ neuron (1 if the neuron exhibits a spike and 0 if it is silent). $\boldsymbol\theta=(\theta_1,\theta_2,\hdots,\theta_N,\theta_{12},\hdots,\theta_{N-1,N})^{\prime}$ is a parameter vector of the Ising model. The second-order parameters $\theta_{ij}$ represent pairwise interactions between neurons. $\psi$ is a log normalization function which serves to ensure the sum of all probabilities equals to 1. The model  in this form is not dependent on time. Hence fitting this model to the data assumes that samples are generated from the same distribution independently at every time step. However, since neuronal activity of in-vivo animals is dynamic \cite{Aertsen89,Vaadia95}, it is necessary to augment the model by allowing $\boldsymbol\theta$ to vary in time. Naively fitting the Ising model at each time step would result in overfitted models unless we had an excessive amount of data. To avoid the issue, we used a sequential Bayesian algorithm to estimate the time-varying parameters. In this framework, we assume the following dynamics for the state $\boldsymbol\theta_t$:
\begin{equation}
\boldsymbol\theta_t=\boldsymbol\theta_{t-1}+\boldsymbol\xi_t(\mathbf{Q}),
\end{equation}
for $t=2,\ldots,T$. At the first time bin, we consider a Gaussian prior defined by $\boldsymbol\theta_1 \sim \mathcal{N}(\boldsymbol\mu,\boldsymbol\Sigma)$. $\xi_t(\mathbf{Q})$ is a 0-mean Gaussian noise added at every time step to obtain stochastic dynamics. The covariance matrix of the noise is given by $\mathbf{Q}=\lambda^{-1}\mathbf{I}$, where $\lambda$ is the precision and $\mathbf{I}$ is the identity matrix. Under the principle of maximizing the marginal log likelihood, it is possible to obtain the optimal set of hyperparameters $\mathbf{w}=\left[\boldsymbol\mu,\boldsymbol\Sigma,\mathbf{Q} \right]$ by using the expectation-maximization (EM) algorithm. The EM algorithm also provides the posterior density of the state $\boldsymbol\theta_t$ for all time bins given the observed data, namely a distribution of the underlying process $\boldsymbol\theta_{1:T}$:
\begin{equation}
p(\boldsymbol\theta_{1:T}\vert\mathbf{x}_{1:T},\mathbf{w})=\frac{p(\mathbf{x}_{1:T}\vert\boldsymbol\theta_{1:T})p(\boldsymbol\theta_{1:T}\vert\mathbf{w})}{p(\mathbf{x}_{1:T}\vert\mathbf{w})}.
\label{eq:filter}
\end{equation}
This posterior density is approximated by a Gaussian distribution. The uncertainty for the parameter estimation is then assessed by its covariance matrix. See \cite{Shimazaki2009,Shimazaki2012} for details of the EM algorithm and sequential Bayes method.

We randomly selected 3 populations of 12 neurons for each monkey (a total of 9 populations). A separate dynamic state-space Ising model was fitted for each stimulus orientation for each population. To quantitatively determine the effect of pairwise interactions, we compared models fitted to the original data with models fitted to surrogate data (surrogate models). The surrogate data was constructed by randomizing the order of the trials for every neuron. This shuffling of the data destroys correlations between neurons, but preserves their spike rate dynamics. Thus, by comparing original models with surrogate models, we can determine if the observed interactions have significant contributions.

\subsection{Macroscopic Properties of the Dynamic Ising Model} After fitting the models, we can investigate the dynamics of the macroscopic properties of the populations during the stimulus exposition. First, the entropy, or the expectation of the information content, is given by
\begin{equation}
S_{pair}(t) = \langle-\log p(\mathbf{x}\vert\boldsymbol\theta_t)\rangle_{\mathbf{x}\vert\boldsymbol\theta_t},
\end{equation}
where the brackets indicate the expectation by the observation density $p(\mathbf{x}|\boldsymbol\theta_t)$. The model containing $N$ binary elements with the maximal entropy is the uniform model where each element has a firing rate of 0.5. Such a model has entropy $S_0=N\log{2}$. By adding information about the firing rates, we reduce the entropy by constraining the model. We call $S_{ind}$ the entropy of the Ising model projected to an independent model which considers the firing rates of individual neurons, but does not exhibit any correlation ($\theta_{ij}=0$ for $i<j$). Considering pairwise interactions also decreases entropy as it constrains the model even more ($S_{pair}$). To assess the contribution of the pairwise interactions in the information content of the population activity, we can compute the fraction of the entropy reduction caused by considering pairwise interactions in the model as   
\begin{equation}
\gamma(t) = \frac{S_{ind}(t)-S_{pair}(t)}{S_0-S_{pair}(t)}
\label{entropy_fraction}.
\end{equation}
 Next, the probability that all neurons are silent, i.e., the sparseness, is given by
\begin{equation}
p_{silence}(t)=p(0,0,\hdots,0\vert\boldsymbol\theta_t)=\exp\left[-\psi(\boldsymbol\theta_t)\right].
\label{eq:silence}
\end{equation}
Finally, the fluctuation of a population, or heat capacity, is the variance of the information content. It represents the sensitivity of the model to changes in the state vector $\boldsymbol\theta_t$. It is defined as
\begin{equation}
C(t)=\langle\{-\log p(\mathbf{x}\vert\boldsymbol\theta_t)\}^2\rangle_{\mathbf{x}\vert\boldsymbol\theta_t}
-\{\langle-\log p(\mathbf{x}\vert\boldsymbol\theta_t)\rangle_{\mathbf{x}\vert\boldsymbol\theta_t}\}^2.
\label{eq:heat_capacity}
\end{equation}

\subsection{Assessment of Stimulus Coding} We also assessed the contribution of pairwise interactions in encoding the stimulus orientation by comparing the neural responses to different stimulus orientations. To do so, we compared the parameters of Ising models fitted to the neural activity of monkeys exposed to gratings of different orientations. Since the EM algorithm provides the posterior density of the state vector approximated as a Gaussian, we computed the Bhattacharyya distance between the posterior densities. The Bhattacharyya distance between two Gaussians $\mathcal{N}(\boldsymbol\mu_1,\boldsymbol\Sigma_1)$ and $ \mathcal{N}(\boldsymbol\mu_2,\boldsymbol\Sigma_2)$ is given as

\begin{equation}
\label{eq:distance}
D_B=\frac{1}{8}(\boldsymbol\mu_1-\boldsymbol\mu_2)^{\prime}\boldsymbol\Sigma^{-1}(\boldsymbol\mu_1-\boldsymbol\mu_2)+\frac{1}{2}\log \left(\frac{\det\boldsymbol\Sigma}{\sqrt{\det\boldsymbol\Sigma_1\det\boldsymbol\Sigma_2}}\right),
\end{equation}
where $\boldsymbol\Sigma=\frac{\boldsymbol\Sigma_1+\boldsymbol\Sigma_2}{2}$. We computed this distance at each time bin. The difference in neural responses is quantified by summing the distances at every time bin.

\section{Results}

\subsection{Contributions of Interactions to Macroscopic Network Properties} Using the time-dependent Ising model, we analyzed the population activity of monkey V1 neurons exposed to an oriented grating stimulus. In total, 9 populations (3 per monkey) were separately analyzed. Results with time bins of 10 ms will be shown here, but we found similar results with 5 and 20 ms. The Bayesian algorithm used to fit the model gives the Gaussian-approximated posterior density of the parameters of the Ising model (Eq.~\ref{eq:ising_model}) given the data, which allows us to obtain the most probable state, or a maximum a posteriori (MAP) estimate, and the credible interval of the estimate (Eq.~\ref{eq:filter}). The fitted model can be used to calculate dynamics of macroscopic properties of the neural populations.

\begin{figure}[t]
\includegraphics[width=1\textwidth,height=57mm]{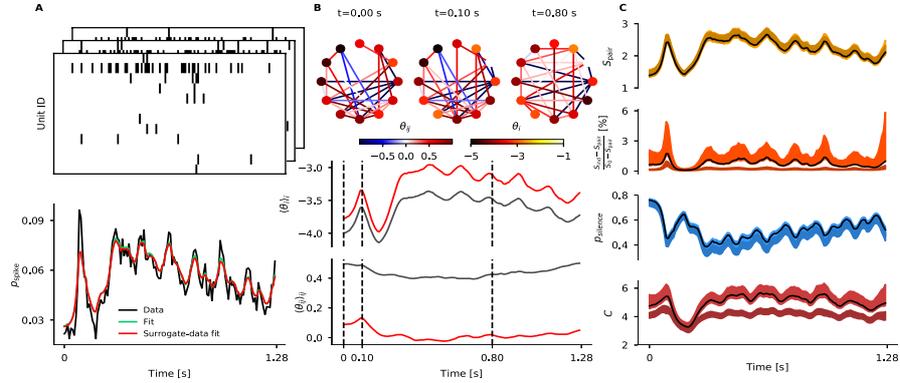}
\caption{\textbf{A} (\textit{Top}) Simultaneous activity of 12 neurons with a 10 ms bin size at exemplary trials from the total 200 trials. The stimulus ($\phi$ = \ang{300}) is presented from 0 s to 1.28 s (\textit{Bottom}) Empirical and estimated population spiking probability. \textbf{B} (\textit{Top}) Snapshots of the estimated parameters of the Ising model. The color of the nodes and edges represent $\theta_{i}$ and $\theta_{ij}$. (\textit{Bottom}) First-order time-dependent parameters averaged over neurons (\textit{Top panel, black line}), and second-order parameters averaged over all pairs (\textit{Bottom panel, black line}). \textit{Red lines} correspond to trial-shuffled data. Vertical dashed bars correspond to the timings of the snapshots. \textbf{C} (\textit{From top to bottom}) Estimates of the entropy, entropy reduction due to interactions, sparseness, and heat capacity (\textit{Black lines}) and their 90\% credible intervals (\textit{Pale shaded area}). The dark shaded areas correspond to the 90\% credible intervals obtained for trial-shuffled data. 
}
\label{fig1}
\end{figure}

Fig.~\ref{fig1} shows results from one exemplary population of 12 neurons. The spike data was recorded 200 times from the same neurons under the same stimulus conditions (here the stimulus orientation ($\phi$) is \ang{300}). Fig.~\ref{fig1}A Top shows the time-steps (x-axis) during which each neuron of the population (y-axis) exhibited spikes (black marks) for 3 exemplary trials. The average spike rate of this population transiently increased about 60 milliseconds after the stimulus onset, as expected for V1 neurons \cite{thorpe2001seeking}, and exhibited oscillatory activity in response to the grating stimulus (Fig.~\ref{fig1}A Bottom). It is thus important to take the rate dynamics into account to assess the correlations among neurons. The state-space Ising model adequately estimated the rate dynamics. Similar rates were observed for other stimulus orientations and populations.

Snapshots of the estimated parameters of the Ising model are shown in Fig.~\ref{fig1}B Top. The colors of the nodes and edges show the values of the MAP estimates for the first-order parameters ($\theta_i$) and the second-order parameters ($\theta_{ij}$), respectively. Only significant edges are shown, for which the value 0 is outside of the 95\% credible interval of the posterior density. The average MAP estimates of the first and second order parameters of the dynamic Ising model can be observed in Fig.~\ref{fig1}B Bottom (black lines). While the first order parameters follow a similar dynamic to that of the firing rate, the interaction parameters only vary on a small scale and with no apparent oscillation. 

Macroscopic measures of the population are shown in Fig.~\ref{fig1}C. The black lines are computed from the MAP estimates of the model parameters. The pale shaded areas correspond to the interval between the 5\% and 95\% quantiles. To compute the quantiles, we sampled $\boldsymbol\theta_{t}$ at every time bin 1000 times from the posterior and computed the macroscopic properties for every sample. 

First, the entropy of the pairwise model ($S_{pair}$) quantifies the information that the population can carry using rates and pairwise interactions. That is to say, the effective number of spiking patterns they can represent is $2^{\frac{1}{log2} S_{pair}}$. Typically, the entropy increases as the probability of spiking increases toward 0.5 (maximum entropy for independent neurons). However, the population activity is  constrained by pairwise interactions, which leads to a reduction of the entropy from the independent assumption. In order to examine the contribution of pairwise interactions in the entropy, we computed the fraction of the entropy reduction caused by considering pairwise interactions in the model $\gamma(t)$ (Eq.~\ref{entropy_fraction}). We found that the pairwise interactions explain approximately $2\%$ of the difference of entropy between the pairwise Ising model and the uniform distribution. 

\begin{figure}
\centering
\includegraphics[width=0.8\textwidth,height=57mm]{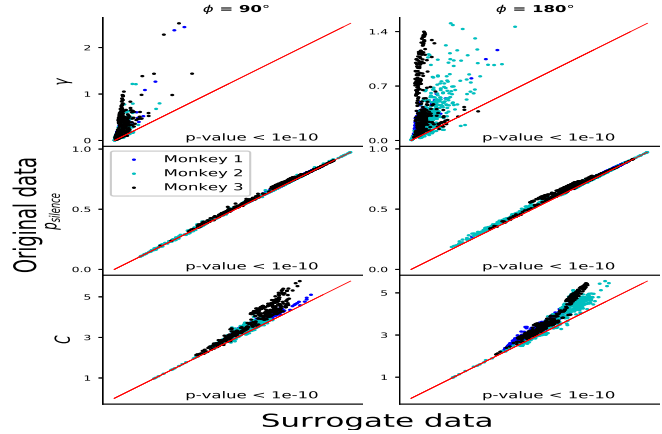}
\caption{Comparison between the properties obtained with original data (\textit{y-axis}) and trial-shuffled data (\textit{x-axis}) from 3 monkeys exposed to gratings at \ang{90} and \ang{180}.
(\textit{From top to bottom}) Entropy reduction due to interactions, sparseness, and heat capacity. 
}
\label{comp}
\end{figure}

To determine if the observed fraction of entropy $\gamma$ is significant, we fitted Ising models to surrogate data. In the surrogate data, the order of the experimental trials was randomized for every neuron in order to destroy interactions. Results are reported by the red lines and the dark shaded areas. The average of the $\theta_{ij}$ parameters (Fig.~\ref{fig1}B Bottom) and the $\gamma$ of the surrogate model being close to 0 confirms that shuffling the trials effectively removed pairwise interactions. The surrogate model also accurately estimated the firing rates (see Fig.~\ref{fig1}A Bottom). By comparing the $\gamma$ obtained with the original and surrogate data, we conclude that there are significant pairwise interactions during the stimulus presentation, as the credible intervals do not coincide. 

We then examined how the pairwise interactions contribute to other macroscopic quantities of the population. The third panel of Fig.~\ref{fig1}C displays the sparseness, i.e., the probability of an all silent pattern (Eq.~\ref{eq:silence}), and the fourth panel displays the heat capacity (Eq.~\ref{eq:heat_capacity}). In this example, the heat capacity was clearly greater for the original model, indicating that interactions of neurons significantly contribute to increasing the sensitivity of the population activity. However, the effect on sparseness may not be obvious. To clarify, next we examined these macroscopic values using all populations. 

Fig.~\ref{comp} compares the macroscopic properties computed with the original and surrogate data. Data points for every populations at every time step are displayed on this figure. As expected, the original models had a bigger $\gamma$. This is because interactions were destroyed in the surrogate data. The original models also displayed significantly bigger sparseness and heat capacity (signed-rank tests). Only results at $\phi=\ang{90}$ and $\phi=\ang{180}$ are shown, but the sparseness and fluctuation were significantly greater for the original data for all orientations.

\begin{figure}[t]
\centering
\includegraphics[width=\textwidth,height=57mm]{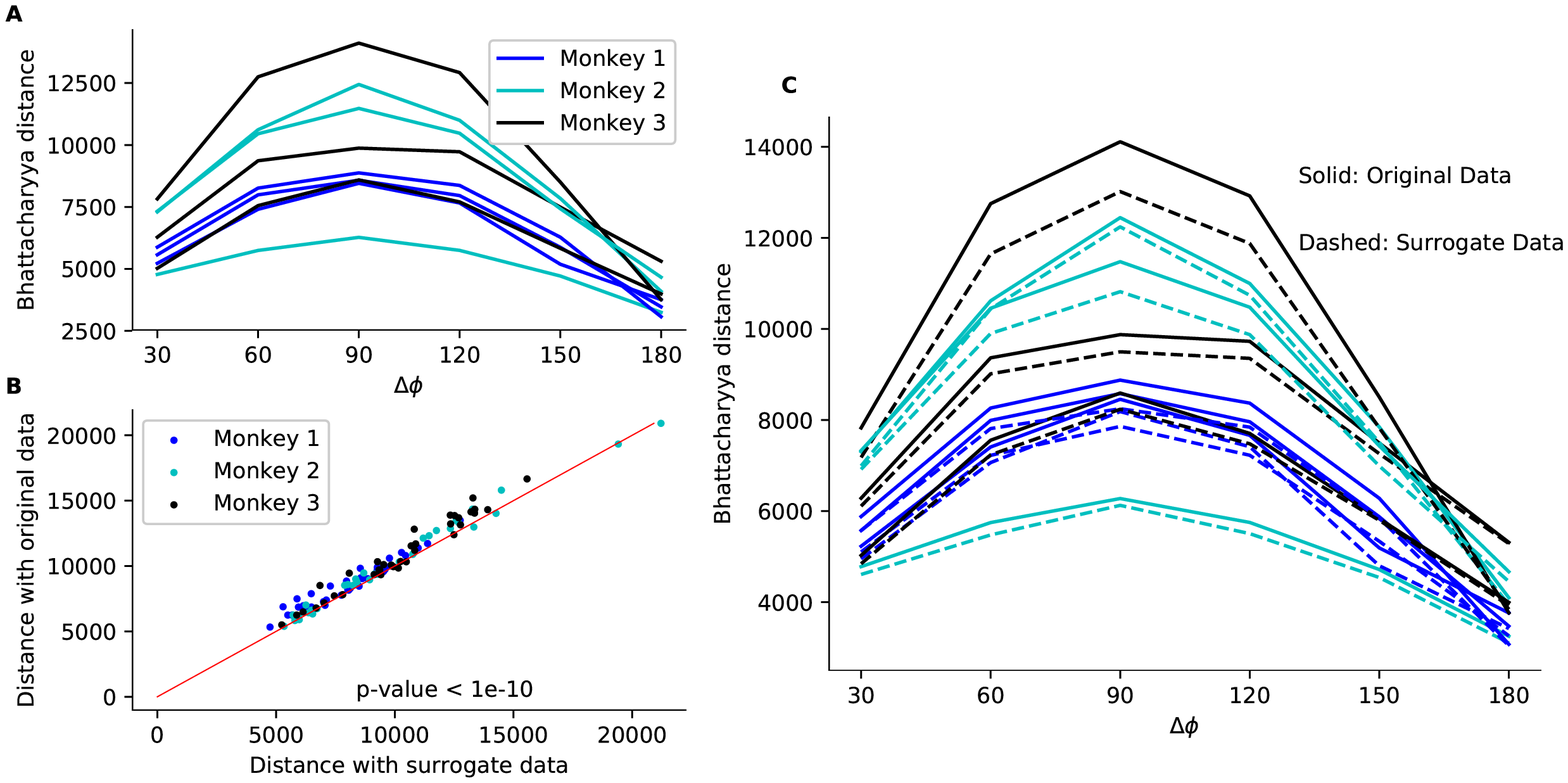}
\caption{\textbf{A} Average Bhattacharyya distance between distributions of parameters of Ising models fitted to monkey V1 neural activity when exposed to sinusoidal gratings at different orientations with respect to the difference of orientation ($\Delta\phi$). \textbf{B} Comparison of the Bhattacharyya distances obtained with models fitted to original data (\textit{y-axis}) and trial-shuffled data (\textit{x-axis}) for all pairs of stimulus orientations separated by \ang{90}. \textbf{C} Average Bhattacharyya distances with respect to the difference of orientation for original data (\textit{full lines}) and trial-shuffled data (\textit{dashed lines}).}
\label{bdistance}
\end{figure}

\subsection{Differences in Neural Responses Caused by Different Stimuli} Next we compared models obtained for different stimulus orientations. This should give an idea of how differently the neurons respond to different gratings orientations. To measure the difference, we computed the Bhattacharyya distance (Eq.~\ref{eq:distance}) between the estimated distributions of the Ising model parameters fitted to neural activity of monkeys when exposed to two different stimulus orientations. For a given population, we summed the distances between the Ising models computed at each time step for all possible pairs of stimulus orientations. We represent the summed Bhattacharyya distance as a function of the difference between stimulus orientations ($\Delta\phi$). We repeated the computations for all populations (Fig.\ref{bdistance}A). The distances exhibited a maximum at $\Delta\phi=\ang{90}$ and a minimum at $\Delta\phi=180$. This means that the population activities were maximally different for two perpendicular stimuli. The stimuli separated by \ang{180} have the same spacial alignment, but their gratings move in opposite directions (e.g., right to left or left to right). Hence the minimum distances at \ang{180} indicate less sensitivity of the population activity to the direction of the stimulus gratings, which is expected from a population of simple cells.

In order to examine contributions of pairwise interactions to the Bhattacharyya distances, the above procedure was also done for surrogate data. Fig.~\ref{bdistance}B shows a comparison of the distances obtained at $\Delta\phi=\ang{90}$ for original and surrogate data. The distances between original models are significantly greater (signed-rank test). Significant increases of the distances were found for all $\Delta\phi$. Fig.~\ref{bdistance}C displays the Bhattacharyya distances computed from original and surrogate models for all $\Delta\phi$. The distances from original data (full lines) are consistently larger than their corresponding surrogate result (dashed line). From this, we conclude that the interactions contributed to increasing the differences in neural activity when the monkeys are exposed to different stimuli. We repeated the same analysis with the Kullback-Leibler divergence between the estimated observation models at different orientations and reached the same conclusions.

\begin{figure}[t]
\includegraphics[width=1\textwidth,height=57mm]{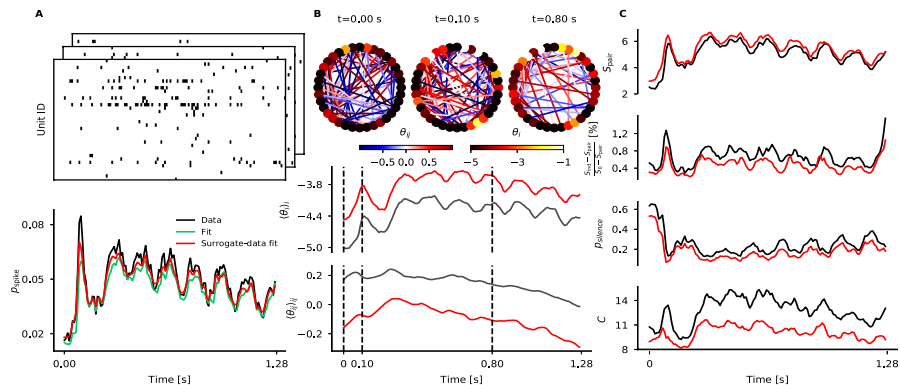}
\caption{\textbf{A} (\textit{Top}) Simultaneous activity of 36 neurons with a 10 ms bin size at exemplary trials from the total 200 trials. The stimulus ($\phi$ = \ang{300}) is presented from 0 s to 1.28 s (\textit{Bottom}) Empirical and estimated population spiking probability. \textbf{B} (\textit{Top}) Snapshots of the estimated parameters of the Ising model. The color of the nodes and edges represent $\theta_{i}$ and $\theta_{ij}$. (\textit{Bottom}) First-order time-dependent parameters averaged over neurons (\textit{Top panel}), and second-order parameters averaged over all pairs (\textit{Bottom panel}). Vertical dashed bars correspond to the timings of the snapshots. \textbf{C} (\textit{From top to bottom}) Estimates of  the entropy, entropy reduction due to interactions, sparseness, and heat capacity. \textit{Black and red lines} correspond to original and trial-shuffled data. 
}
\label{fig4}
\end{figure}

\section{Discussion}
We found a significant contribution of pairwise interactions to stimulus encoding. Since the neural population activity is more different with respect to the stimulus in the presence of pairwise interactions, the interactions should improve the decoding of stimulus information. However, we found a small percentage of entropy due to pairwise interactions ($\sim 2\%$). While this may be caused by the small number of neurons or by the use of a simple Gabor artificial stimuli instead of correlated natural stimuli \cite{Ganmor2011}, considering the firing rate dynamics might have successfully removed spurious correlations. Previous analyses based on the stationary model may suffer from the spurious spike correlations caused by rate covariations. Our analysis reveals that neurons exhibit near-independent activity during stimulus presentation. This result is consistent with the efficient use of population activity expected from the efficient coding hypothesis \cite{barlow1961possible,olshausen1997sparse}.

Donner et al. \cite{Donner2017} introduced approximation methods (pseudo-likelihood combined with TAP or Bethe approximation) to fit the state-space Ising model to larger networks. We used these methods to fit models to 1 population of 36 neurons per monkey (Fig.~\ref{fig4}, the same monkey and stimulus orientations as shown in Fig.~\ref{fig1}). The results were consistent with those obtained in the exact analysis: pairwise interactions had significant contributions to increasing sparseness and sensitivity for all monkeys and orientations (signed-rank test). We chose to provide the results of an analysis without the approximations, but our conclusions regarding sparseness and heat capacity are robust to the network size. 

\section{Conclusion}
The neural interactions significantly contributed to shaping the activity of monkey V1 neurons when exposed to sinusoidal gratings. Neuron populations present significant sparseness and sensitivity due to the neurons' interactions. Neural activities are organized differently when neurons respond to different stimulus orientations, and this difference is enhanced by the presence of neural interactions. From this result, we expect that the decoding of the stimulus orientation is facilitated by considering pairwise interactions of the neurons. 

\bibliographystyle{splncs03}
\bibliography{references}

\end{document}